\documentclass[oldversion]{aa}
\usepackage{amsmath,graphicx,amssymb}
\usepackage{natbib}
\bibpunct{(}{)}{,}{a}{}{,}

\newcommand{\zav}[1]{\left(#1\right)}
\newcommand{\hzav}[1]{\left[#1\right]}

\newcommand{\azav}[1]{\left|#1\right|}
\newcommand\de{\text{d}}
\newcommand\req{\ensuremath{R_\text{eq}}}
\newcommand\Sun{\odot}%
\newcommand\aalpha{\tilde\alpha}

\begin{document}

\title{Mass and angular momentum loss via decretion disks}

\author{J.~Krti\v{c}ka\inst{1}, S.~P.~Owocki\inst{2} and
G.~Meynet\inst{3}}
\authorrunning{J.~Krti\v{c}ka, S. P. Owocki \& G.~Meynet}


\institute{Department of Theoretical Physics and Astrophysics,
           Masaryk University, CZ-611\,37 Brno, Czech Republic,
           \email{krticka@physics.muni.cz}
           \and
           Bartol Research Institute, University of Delaware, Newark,
           DE 19716, USA
           \and
           Geneva Observatory, CH-1290 Sauverny, Switzerland}

\date{Received}

\abstract{%
We examine the nature and role of mass loss via an equatorial decretion disk
in massive stars with near-critical rotation induced by evolution of
the stellar interior.
In contrast to the usual stellar wind mass loss set by exterior driving
from the stellar luminosity, such decretion-disk mass loss stems from 
the angular momentum loss needed to keep the star near and below critical 
rotation, given the interior evolution and decline in the star's moment
of inertia.
Because the specific angular momentum in a Keplerian disk
increases with the square root of the radius,
the decretion mass loss associated with a
required level of angular momentum loss depends crucially on the outer radius
for viscous coupling of the disk, and can be significantly less than
the spherical, wind-like mass loss commonly 
assumed in evolutionary calculations.
We discuss the physical processes that affect the outer disk
radius, including thermal disk outflow, and ablation of the 
disk material via a line-driven wind induced by the star's radiation. 
We present parameterized scaling laws for taking account of decretion-disk 
mass loss in stellar evolution codes, including how these are affected by
metallicity, or by presence within a close binary and/or a dense
cluster. Effects similar to those discussed here should also be present
in accretion disks during star formation, and may play an important
role in shaping the distribution of rotation speeds on the ZAMS.}

\keywords {stars: mass-loss -- stars: evolution -- stars: rotation --
hydrodynamics}

\maketitle

\section{Introduction}

Classical models of stellar evolution focus on the dominant role of 
various stages of nuclear burning in the stellar core.
But in recent years it has become clear that stellar evolution,
particularly for more massive stars, can also be profoundly influenced
by the loss of mass and angular momentum from the stellar envelope and
surface \citep{mm08}.
In cool, low-mass stars like the sun, mass loss through thermal expansion 
of a coronal wind occurs at too-low a rate to have a direct
effect on its mass evolution; 
nonetheless the moment arm provided by the coronal
magnetic field means the associated wind angular momentum loss can
substantially spin down the star's rotation as it ages through its
multi-Gyr life on the main sequence.
Except in close binary systems, the rotation speeds of cool, low-mass 
stars are thus found to decline with age, from up to $\sim$100~km s$^{-1}$ near the
ZAMS to a few km s$^{-1}$ for middle-age stars like the sun.

By contrast, in hotter, more massive stars the role and nature of mass and 
angular momentum loss can be much more direct and profound, even over 
their much shorter, multi-Myr lifetimes.
While some specific high-mass stars appear
to have been spun down by strongly magnetized stellar winds
(e.g HD 191612, \citealt{donetal06}, or HD 37776, \citealt{mikbra}),
most massive stars are comparitively rapid rotators, with typical
speeds more than 100~km s$^{-1}$, and in many stars, e.g.\ the Be stars, even 
approaching the {\em critical} rotation rate, at which the 
centrifugal acceleration at the equatorial surface balances Newtonian gravity
\citep{how04, toh, how07}.

Indeed, models of the
MS evolution of rotating massive stars show that, at the surface, the
velocity approaches the critical velocity.
This results from the transport of angular momentum from the contracting,
faster rotating inner convective core  to the expanding, slowed down
radiative envelope
\citep{mee}.
In stars with moderately rapid initial rotation, and with only moderate
angular momentum loss from a stellar wind, this spinup from internal
evolution can even bring the star to critical rotation
\citep{memb07}.
Since any further increase in rotation rate is not dynamically allowed,
the further contraction of the interior must then be balanced 
by a net loss of angular momentum through an induced mass loss.
In previous evolutionary models, the required level of mass loss has
typically been estimated by assuming its removal occurs from
spherical shells at the stellar surface \citep{mee}.

This paper examines the physically more
plausible scenario that such mass loss occurs through an {\em equatorial,
viscous decretion disk} \citep{los91}.
Such decretion disk models have been extensively applied to analyzing the
rapidly (and possibly near-critically) rotating Be stars, which show
characteristic Balmer emission thought to originate in geometrically
thin, warm, gaseous disks in Keplerian orbit near the equatorial plane
of the parent star
\citep{porriv, carbjo}.
But until now there hasn't been much consideration of the role such viscous 
decretion disks might play in the rotational and mass loss {\em evolution} 
of massive stars in general.

As detailed below, a key point of the analysis here is that, per unit 
mass, the angular momentum loss from such a decretion disk can greatly
exceed that from a stellar wind outflow.
Whereas the angular momentum loss of a nonmagnetized wind is fixed
around the transonic point very near the stellar surface,
the viscous coupling in a decretion disk can transport angular
momentum outward to some outer disk radius
$R_\text{out}$, where the specific angular momentum is a factor
$\sqrt{R_\text{out}/R_\text{eq}}$ higher than at the equatorial surface.
For disks with an extended outer radius $R_\text{out} \gg R_\text{eq}$, the
angular momentum loss required by the interior evolution can then be
achieved with a much lower net mass loss than in the wind-like, 
spherical ejection assumed in previous evolution models.

For a given angular momentum shedding mandated by interior evolution,
quantifying the associated disk mass loss thus requires determining the 
disk outer radius.
For example, in binary systems, this would likely be limited by tidal 
interactions with the companion, and so scale with the binary
separation
\citep{ok02}.
But in single stars, the processes limiting this outer radius are less
apparent.
Here we explore two specific mechanisms that can limit the
angular momentum loss and/or outer radius of a disk, namely 
thermal expansion into supersonic flow at some outer radius, and 
radiative ablation of the inner disk from the bright central star.
For each case, we derive simple scaling rules for the required disk
mass loss as a function of assumed stellar and wind parameters, given 
the level of interior-mandated angular momentum loss.

The organization for the remainder of this paper is as follows:
Sect.~\ref{secana} presents simple analytical relations for how the presence
of a disk affects the mass loss at the critical limit. 
Sect.~\ref{secnu} develops set of equations governing structure and kinematics of the
disk, while Sect.~\ref{kaprenum} solves these to derive simple scaling
for how thermal expansion affects the outer disk radius and disk mass
loss.
Sect.~\ref{secabla} discusses the effects of inner-disk
ablation by a line-driven disk wind induced from the illumination
of an optically thick disk
by the centeral star, deriving the associated
ablated mass loss and its effect on the net disk angular momentum and mass loss.
Sect.~\ref{secevol} gives a synthesis of the different cases discussed here and
offers a specific recipe for incorporating disk mass loss rates into stellar
evolution codes. 
Sect.~\ref{secot} discusses some complementary points (e.g.\ viscous
decoupling, tidal effects of nearby stars, reduced metallicity, etc.), 
while Sect.~\ref{seccon} concludes with a brief
summary of the main results obtained in this work.

\section{Basic analytic scaling for disk mass loss}
\label{secana}
Let us begin by deriving some simple analytic expressions for the effect of the disk viscous
coupling on the disk mass-loss rate.

Assuming a star that rotates as a rigid body, the magnitude of stellar angular momentum $J$
is given by the product of the stellar moment of inertia $I$ and the rotation angular 
frequency $\Omega$, $J=I\Omega$.
During stellar evolution, the time rate of change of angular momentum depends on the
changes in moment of inertia and rotation frequency,
\begin{equation}
\dot J = \dot I \Omega + I \dot\Omega.
\end{equation}
If, for example, the moment of inertia declines at a rate $\dot I$, and the
change of the angular momentum through any wind, etc. is negligible, i.e. $\dot
J = 0$, then the star has to spin up at a rate given by
\begin{equation}
\frac{\dot\Omega}{\Omega}=-\frac{\dot I}{I}.
\end{equation}
However, once the star reaches the critical rotation frequency $\Omega =
\Omega_\text{crit} \equiv \sqrt{GM/\req^3}$ (where $M$ is the stellar mass 
and  \req\ is the equatorial radius when the star is rotating at the critical limit), this spin-up has to end ($\dot\Omega=0$), 
requiring instead a shedding of angular momentum given by
\begin{equation}
\dot J = \dot I \Omega_\text{crit}.
\end{equation}

If we assume this occurs purely through mass loss at a rate $\dot M$ through a
Keplerian decretion disk, the angular momentum loss is set by the
outer radius $R_\text{out}$ of that disk, given by
\begin{equation}
\label{dotjap}
\dot J_\text{K} (R_\text{out}) = \dot M v_\text{K}(R_\text{out}) R_\text{out} =
\dot M \Omega_\text{crit} \req^2 \sqrt{\frac{R_\text{out}}{\req}},
\end{equation}
where the Keplerian velocity\footnote{When the star is rotating at the critical
limit, the critical rotational velocity is equal to the Keplerian velocity at
$\req$, $v_\text{K}(\req)=\Omega_\text{crit}\req$.} is
$v_\text{K}(r)=\sqrt{GM/r}$. Setting $\dot J_\text{K} (R_\text{out})$ equal to the above $\dot J$
required by a moment of inertia change $\dot I$, we find the required mass
loss rate is
\begin{equation}
\label{mdotdisk}
\dot M = \frac{\dot I}{\req^2} \sqrt{\frac{\req}{R_\text{out}}}.
\end{equation}
As $R_\text{out}$ gets larger, note that the required mass loss rate gets smaller.

Eq.~\eqref{mdotdisk} can be compared with the case where mass  decouples in a
spherical shell just at
the surface of the star, i.e., where $R_\text{out}=\req$. In that case the
required mass loss is just
\begin{equation}
\label{dmdtat}
\dot M  = \frac{3}{2}\frac{\dot I}{\req^2}.
\end{equation}
So when a Keplerian disk is present, the mass loss is reduced by a factor
$\frac{3}{2}\sqrt{R_\text{out}/\req}$ with respect to the case with no disk.
If $R_\text{out}$ is small, large mass-loss is necessary to shed the
required amount of angular momentum to keep the rotation frequency at its
critical value. In the opposite case, when $R_\text{out}$ becomes substantial,
only a relatively small mass-loss is necessary to shed the required amount of
angular momentum, implying a significant difference in the mass loss evolution.

\section{Numerical models}
\label{secnu}

\begin{figure*}
\centering
\includegraphics[width=0.45\hsize]{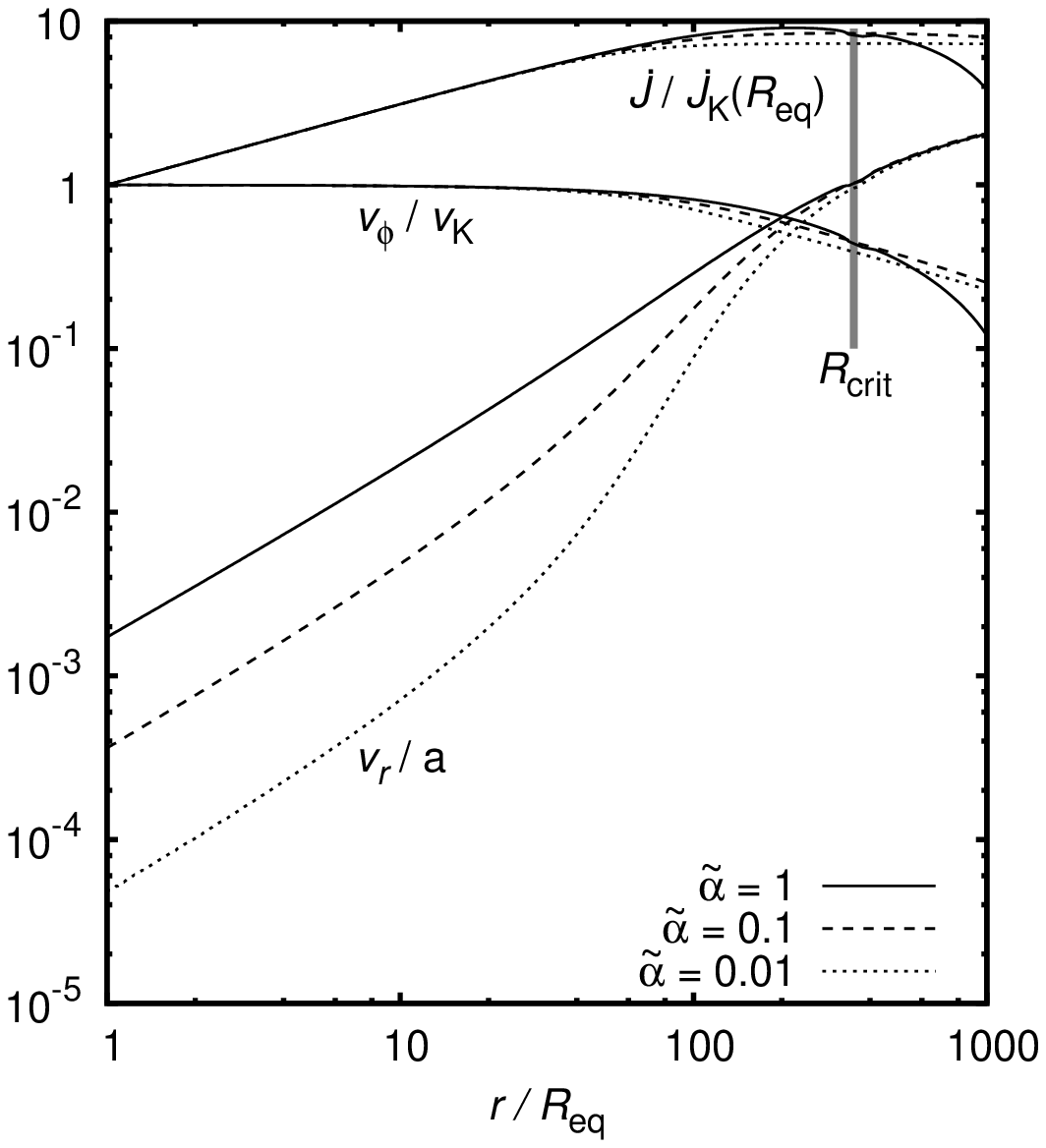}
\includegraphics[width=0.45\hsize]{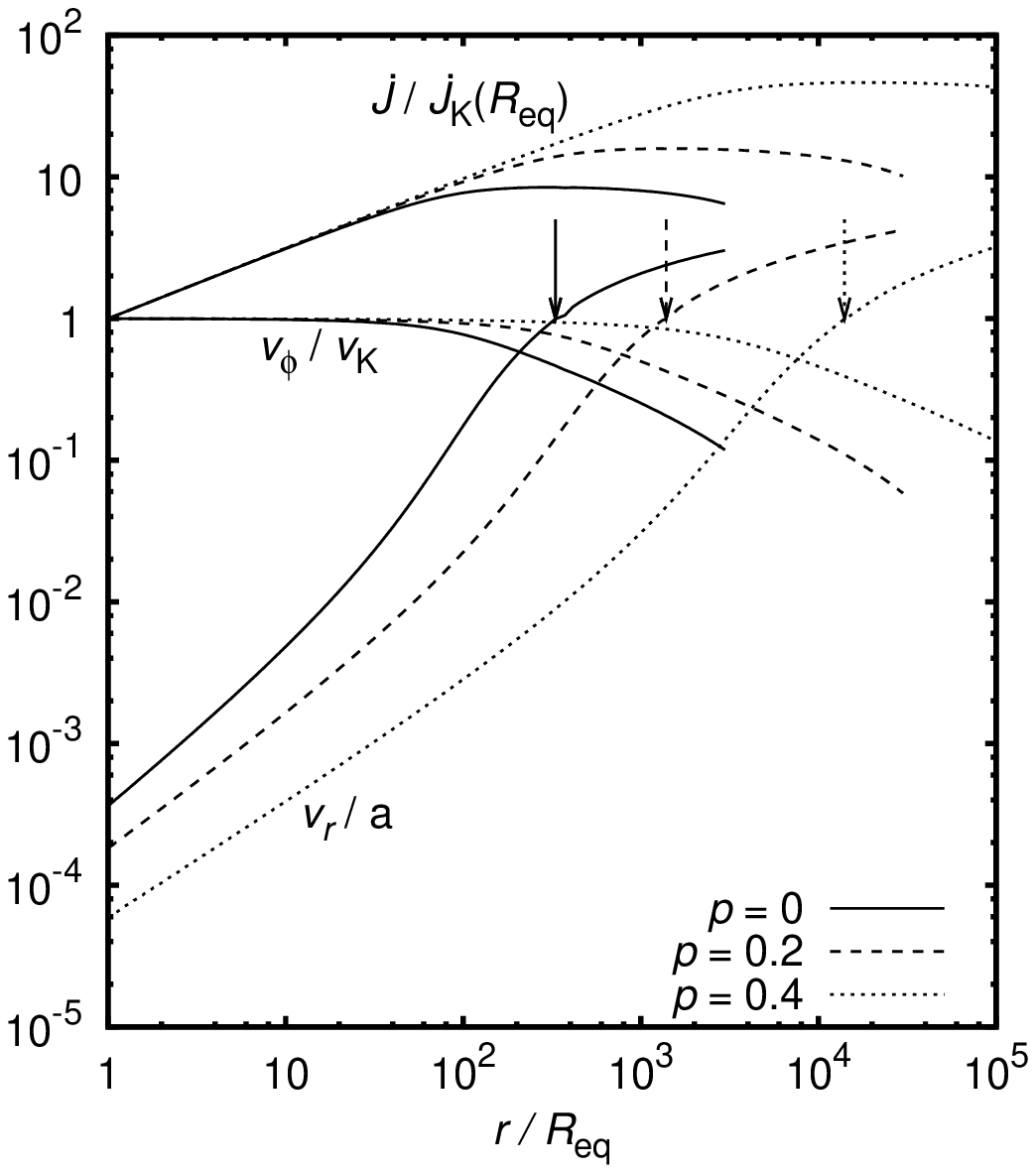}
\caption{The dependence of the radial velocity, 
azimuthal
velocity, and the
angular momentum loss
rate
in units of 
equator release angular momentum loss rate $\dot J_\text{K}
(\req)$ on the radius in a
viscous disk. Left: models of isothermal disk ($p=0$,
$T_0=\frac{1}{2}T_\text{eff}$)
with different viscosity
parameter $\aalpha$. Right: Models with various temperarure profile for fixed
$\aalpha=0.1$ and $T_0=\frac{1}{2}T_\text{eff}$. Arrows denote the location of
critical points.}
\label{500_2_a}
\end{figure*}

To obtain a detailed disc structure, we solve stationary hydrodynamic
equations in cylindrical coordinates integrated over the height above the
equatorial plane $z$ \citep{lman,los91,sapporo,josip}. Assuming axial symmetry, 
the corresponding variables,
i.e., the radial and 
azimuthal
velocities $v_r$, $v_\phi$, and the integrated
disk density $\Sigma=\int_{-\infty}^{\infty}\rho\,\de z$, depend only on radius
$r$. The equation of continuity in
such a case is
\begin{equation}
\label{kontrov}
\frac{1}{r}\frac{\de \zav{r\Sigma v_r}}{\de r}=0.
\end{equation}
The stationary conservation of the $r$ component of momentum gives
\begin{equation}
\label{rrov}
v_r\frac{\de v_r}{\de r}=\frac{v_\phi^2}{r}+g-
\frac{1}{\Sigma}\frac{\de (a^2\Sigma)}{\de r}+\frac{3}{2}\frac{a^2}{r},
\end{equation}
where $g=-GM/r^2$, 
$a^2=kT/(\mu m_\text{H})$, with the temperature assumed to
vary as a power-law in radius, $T=T_0\zav{\req/r}^p$, where $T_0$ and $p$ 
are free parameters,
$\mu$
is the mean molecular weight
(taking $\mu=0.62$),
and $m_\text{H}$ is the mass of a hydrogen atom.
In the equation of conservation of the $\phi$ component of momentum we introduce
the viscosity term \citep{sakura} parametrized via $\aalpha$
\begin{equation}
\label{phirov}
\frac{v_r}{r}\frac{\de \zav{r v_\phi}}{\de r}+
\frac{\aalpha}{r^2\Sigma}\frac{\de}{\de r}\zav{a^2r^2\Sigma}=0,
\end{equation}
and the conservation of the $\theta$ component of momentum gives the hydrostatic
equilibrium density distribution
\begin{equation}
\label{rhodisk}
\rho=\rho_0\exp\zav{-\frac{1}{2}\frac{z^2}{H^2}}, \qquad
H=\frac{a}{v_\text{K}}r.
\end{equation}
Note that the equatorial density $\rho_0$ is related to the 
vertically
integrated disk
density via $\Sigma = \sqrt{2 \pi} \rho_0 H$.
Close to the star, detailed energy-balance models
\citep{milma,carbjo} show the disk is nearly isothermal with
$T_0=\frac{1}{2}T_\text{eff}$ and $p=0$. But to account for the radial
decline of the temperature in the outer regions, we also consider here
models with power law temperature decline, with $p>0$.

The system of equations Eq.~\eqref{kontrov}--\eqref{phirov} has to be
supplemented by appropriate boundary conditions.
For obtaining the value of $v_r$ at the stellar surface $r=R_\text{eq}$ we use
the fact that at the critical point with radius $R_\text{crit}$ given by the
condition
\begin{equation}
\label{krit}
\frac{v_\phi^2}{R_\text{crit}}-\frac{GM}{R_\text{crit}^2}+
\frac{5}{2}\frac{a^2}{R_\text{crit}}-
\left.\frac{\de a^2}{\de r}\right|_{R_\text{crit}}=0
\end{equation}
we should have that $v_r(R_\text{crit})=a$ to ensure the finiteness of the
derivatives at this point \citep[Eqs.~\eqref{kontrov}, \eqref{rrov}, see
also][]{sapporo}.
Thus we
chose $v_r$ at the surface such that at $R_\text{crit}$ we have $v_r=a$. Note
that the radial disk velocity is supersonic above the critical point. The value
of the 
azimuthal
velocity at the stellar surface $v_\phi$ is equal to the
corresponding Keplerian velocity. The system of studied hydrodynamical equations
is invariant for the change of the scale $\Sigma'=\gamma\Sigma$ (where $\gamma$
is constant). Consequently, the equations do not provide any constraint for the
mass-loss rate $\dot M=2\pi r v_r\Sigma$, which in our case is obtained
from the angular momentum loss required by the evolutionary calculations. This
provides the remaining boundary condition for the column density $\Sigma$. Here
we treat the disk mass-loss rate as a free parameter.

For the numerical solution of the system of equations
Eq.~\eqref{kontrov}--\eqref{phirov} we approximate the differentiation at
selected radial grid and use the Newton-Raphson method \citep[e.g.,][]{krtueb}.
The resulting system of linear equations is solved using the numerical package
LAPACK ({\tt http://www.cs.colorado.edu/\~{}lapack}, \citealt{lapack}).

\section{Results of numerical models}
\label{kaprenum}

The general disk properties do not significantly depend on particular stellar
parameters. Nevertheless, to be specific, for a detailed modelling we selected
the stellar parameters roughly corresponding to evolved massive first star
\citep{bezmari,eznula} $T_\text{eff}=30\,000\,\text{K}$, $M=50\,M_\Sun$,
$R=30\,R_\Sun$.

The calculated models for different values of $\aalpha$ are given in
Fig.~\ref{500_2_a}. Close to the star the integration of the momentum equation
Eq.~\eqref{phirov} using the continuity equation Eq.~\eqref{kontrov}
\begin{equation}
\label{momint}
rv_\phi+\frac{\aalpha a^2r}{v_r}=\text{const.}
\end{equation}
gives linear dependence of the radial velocity on radius in isothermal disks
(for $v_r\ll a$), $v_r\sim r$, consequently $\Sigma\sim r^{-2}$ \citep{sapporo}.
Finally, from the momentum equation Eq.~\eqref{rrov}, it follows that close to the
star the disk rotates as Keplerian one, i.e.~$v_\phi\sim r^{-1/2}$. As a result,
the angular momentum loss scales as $\dot J\sim rv_\phi\sim r^{1/2}$, in
accordance with Eq.~\eqref{dotjap}. As the disk accelerates in radial direction,
$v_r$ becomes comparable with $a$ and the term $rv_\phi$ dominates in
Eq.~\eqref{momint}, consequently the disk is momentum conserving close to the
critical point, $rv_\phi=\text{const.}$ (see Fig.~\ref{500_2_a}).

In the supersonic region from the momentum equation Eq.~\eqref{rrov}
follows the logarithmic dependence of the radial velocity on radius,
$v_r^2\sim\ln r$. Consequently, the second term in equation Eq.~\eqref{momint}
rises and as a result of this $v_\phi$ may become even negative.
However, this behaviour is a consequence of adopted
Shakura-Sunyaev viscosity prescription which predicts non-zero torque even for
shear-free disks,
and is likely not applicable in the supersonic region.

A maximum angular momentum loss due to the disk is obtained in the case when the
disk has its outer edge at the radius where $\dot J$ is maximum (see
Fig.~\ref{500_2_a}). Note that this value does not significantly depend on the
assumed viscosity parameter $\aalpha$. An estimate of the maximum angular
momentum loss can be obtained assuming that it is equal to the angular momentum
loss at the critical point. From the numerical models it follows that the
azimuthal
velocity at the critical point is roughly equal to the half of the
Keplerian velocity (see Fig.~\ref{500_2_a}),
\begin{equation}
v_\phi(R_\text{crit}) \approx \frac{1}{2}v_\text{K}(R_\text{crit}).
\end{equation}
In this case the critical point
condition Eq.~\eqref{krit} yields an estimate of the critical point radius
\begin{equation}
\frac{R_\text{crit}}{\req} =
\hzav{\frac{3}{10+4p}\zav{\frac{v_\text{K}(\req)}{a(\req)}}^2}^{\frac{1}{1-p}}
\end{equation}
from which the maximum angular momentum loss via the disk follows
\begin{equation}
\label{mamol}
\dot J_{\aalpha}(\dot M)\approx\frac{1}{2}
\hzav{\frac{3}{10+4p}\zav{\frac{v_\text{K}(\req)}{a(\req)}}^2}^{\frac{1}{2-2p}}
\dot M v_\text{K}(\req) \req.
\end{equation}
In agreement with Fig.~\ref{500_2_a}, comparing the formula Eq.~\eqref{mamol}
with analytical estimate Eq.~\eqref{dotjap} the angular momentum loss is roughly
given by $\frac{1}{2}\dot J_\text{K}(R_\text{crit})$, i.e., it is one half of
the angular momentum loss of the Keplerian disk truncated at the critical point
radius $R_\text{crit}$. The factor $\frac{1}{2}$ comes from the fact that the
disk is not rotating as a Keplerian one at large radii (see Fig.~\ref{500_2_a}).
Hence, the minimum disk mass loss rate required for given moment of inertia
decline is by a factor of about $\zav{{v_\text{K}(\req)}/{a(\req)}}^{1/1-p}$
lower than in the case without a disk.

Note also that adding cooling can substantially increase the critical radius and
thus the disk angular momentum loss. For example, for $p=0.4$ the angular
momentum loss increases by a factor of 10 compared to the isothermal case (see
Fig.~\ref{500_2_a}).

\section{Radiative ablation}
\label{secabla}

As the radiative force may drive large amount of mass out of the hot stars via
line-driven wind \citep[see][for a review]{owopo,puvina} it may also effectively
set the outer disk radius. The radiative force may in this case ablate the
material from the disk and sustain a radiatively driven outflow
\citep{abla,diskra}. In the following we give an estimate of the disk wind
mass-loss rate, which is derived in Appendix.

The disk outflow may in our case resemble the radiation driven winds from
luminous accretion disks \citep{prostod,feshlo,feshlovit,prostod2}. The outflow
in these simulations originates from the whole disk surface. Consequently, part
of the stellar outflow is carried outwards by the disk and part by the disk wind
and the fraction of material carried out by the disk wind increases with radius.
The disk will be in this case truncated at the radius where the material is
carried away entirely by the wind. As the viscous coupling is likely not
maintained in the supersonic wind, only the ablation of the material from the
regions close to the star would decrease the effectiveness of braking.

Let us roughly determine the mass-loss rate of such disk wind. The classical
\citet[hereafter CAK]{cak}
stellar wind mass-loss rate estimate
\begin{equation}
\label{mcak}
\dot M_\text{CAK}=\frac{\alpha}{1-\alpha}\frac{L}{c^2}
\zav{\Gamma\bar Q}^{1/\alpha-1},
\end{equation}
where 
$\bar Q$ and $\alpha$ are line force parameters
\citep[see also][]{gayley},
$L$ is the stellar luminosity,
and
the Eddington parameter $\Gamma=\kappa_\text{e} L /\zav{4\pi GMc}$,
with $\kappa_\text{e}$ beeing the Thomson scattering cross-section per unit of
mass,
can be
rewritten in the term of mass flux from a unit surface,
\begin{equation}
\label{mcakjed}
\dot m=\frac{\alpha}{1-\alpha}\frac{\tilde F}{c^2}
\zav{\frac{\kappa_\text{e} \tilde F\bar Q}{c\tilde g}}^{1/\alpha-1},
\end{equation}
where $\tilde F$ is the driving flux and $\tilde g$ is local gravitational
acceleration. The radiative energy impinging the unit of surface parallel to the
direction to the star is from geometrical reasons proportinal to $FR/r$, where
$R$ is the polar radius, and
$F$ is the radiative flux
at radius $r$. Assuming that
the disk is optically thick (see Sect.~\ref{apdisopdep}), and
all incident radiation is directed upward, we can
roughly estimate
\begin{equation}
\label{fluxjed}
\tilde F=\frac{R}{r}F.
\end{equation}
Taking $\tilde g=GM/r^2$, the total disk wind mass-loss rate is then given by an
integral of the mass-loss
rate per unit of the disk surface $\dot m$ between the equatorial radius
$R_\text{eq}$ and the outer disk radius $R_\text{out}$
\begin{equation}
\label{dmdtdefjed}
\dot M_\text{dw}(R_\text{out})=2\times2\pi
\int_{R_\text{eq}}^{R_\text{out}}\dot m r\,\de r,
\end{equation}
where factor of 2 in Eq.~\eqref{dmdtdefjed} comes from the fact that the wind
originates from both sides of the disk. Inserting the mass flux estimate
Eq.~\eqref{mcakjed} and \eqref{fluxjed} we derive
\begin{equation}
\label{dmdtjed}
\dot M_\text{dw}(R_\text{out})=\dot M_\text{CAK}P_1\zav{\frac{R_\text{out}}{R}},
\end{equation}
where 
(using substitution $x=r/R$)
\begin{equation}
\label{pjed}
P_\ell(x_\text{out})=\int_{3/2}^{x_\text{out}}x^{-1/\alpha-\ell}\de x=
\frac{\zav{\frac{3}{2}}^{1-\ell-\frac{1}{\alpha}}-x_\text{out}^{1-\ell-\frac{1}{\alpha}}}{\frac{1}{\alpha}+\ell-1}.
\end{equation}

Assuming the disk wind is not viscously coupled to the disk, the total angular
momentum loss rate via the disk wind is
\begin{equation}
\label{jobjed}
\dot J_\text{dw}(R_\text{out})=2\times2\pi\int_{R_\text{eq}}^{R_\text{out}}
\dot mv_\phi r^2\, \de r.
\end{equation}
As the disk wind originates mainly
from the regions close to the star (with $r/R\lesssim10$, see
Fig.~\ref{palfaobr}), where the
azimuthal
velocity is roughly equal to the Keplerian one (see
Fig.~\ref{500_2_a}), we can assume $v_\phi\approx v_\text{K}(r)$ in
Eq.~\eqref{jobjed} and consequently the disk wind
angular momentum loss rate
\begin{equation}
\label{dotjjed}
\dot J_\text{dw}(R_\text{out})=R\,v_\text{K}(R)\,P_{\frac{1}{2}}
\zav{\frac{R_\text{out}}{R}}\dot M_\text{CAK}
\end{equation}
is by a factor of $P_{\frac{1}{2}}(R_\text{out}/R)$ larger than the angular
momentum loss due to the CAK wind launched from equator of hypothetical
critically rotating spherical star with radius $R$.

A more detailed calculation (see Appendix~\ref{secablaap}) gives a more
complicated form of $P_\ell(x_\text{out})$ via Eq.~\eqref{paprox}
\begin{equation}
\label{paproxte}
P_\ell(x_\text{out})=\frac{2\pi^{-\frac{1}{\alpha}}
3^{\frac{1}{2\alpha}-\frac{3}{2}}}{\frac{1}{\alpha}+\ell-1}
\hzav{\zav{\frac{3}{2}}^{1-\ell-\frac{1}{\alpha}}-x_\text{out}^{1-\ell-\frac{1}{\alpha}}},
\end{equation}
which shall be used in Eqs.~\eqref{dmdtjed}, \eqref{dotjjed}
instead of Eq.~\eqref{pjed}.

For an infinite disk ($R_\text{out}\rightarrow\infty$) we derive from
Eq.~\eqref{paproxte} maximum disk wind mass-loss rate
\begin{equation}
\label{mdotmax}
\dot M_\text{dw}(\infty)=2^{1+\frac{1}{\alpha}}\pi^{-\frac{1}{\alpha}}
3^{-\frac{1}{2\alpha}-\frac{3}{2}}\alpha
\dot M_\text{CAK},
\end{equation}
and maximum angular momentum loss rate as
\begin{equation}
\label{jdotmax}
\dot J_\text{dw}(\infty)=\frac{2^{\frac{3}{2}+\frac{1}{\alpha}}
\pi^{-\frac{1}{\alpha}}
3^{-\frac{1}{2\alpha}-1}}{2-\alpha}\alpha
R\,v_\text{K}(R)\,\dot M_\text{CAK}.
\end{equation}
For a typical value of $\alpha\approx0.6$ \citep{nlteii} the maximal disk wind
mass-loss rate is relatively low, just about $1/25$ of the CAK stellar wind
mass-loss rate.

\section{Mass loss of the star-disk system at the critical limit}
\label{secevol}

The structure of the decretion disk and the radiatively driven wind blowing from
its surface depends on the value of the angular momentum loss $\dot J$ needed to
keep the stellar rotation at or below the critical rate and on the magnitude of
the radiative force. If the angular momentum loss is small, then the disk could
be blown away by the radiative force already very close to the star. In the
opposite case, if the angular momentum loss is large, then the mass carried away
by the disk wind is negligible. 

In the intermediate case the mass and angular momentum will be carried partly by
the disk and partly by the disk wind. 
When the star has to lose angular momentum at a rate $\dot J$ to keep at most
the critical rotation, the angular momentum will be carried by the
stellar wind ($\dot J_\text{w}$), by the disk wind ($\dot
J_\text{dw}$), and by the disk itself ($\dot J_{\aalpha})$,
\begin{equation}
\label{jecka}
\dot J= \dot J_\text{w} + \dot J_\text{dw} + \dot J_{\aalpha}(\dot M_\text{d}).
\end{equation}
For the calculation of total disk mass-loss rate the following procedure
could be used.

For a given stellar and line-force parameters ($\bar Q$ and $\alpha$)
the maximum disk wind angular momentum loss 
$\dot J_\text{dw}(\infty)$
corresponding to infinite disk $R_\text{out}\rightarrow\infty$ 
can be calculated using Eqs.~\eqref{jdotmax}.
If the net angular momentum loss that should be carried away by
the disk outflow $\dot J-\dot J_\text{w}$ is lower than the maximum one, $\dot
J-\dot J_\text{w}<\dot J_\text{dw}(\infty)$, then the disk will be completely
ablated by the radiation at the radius $R_\text{out}$ given
(from Eq.~\eqref{jecka} for $\dot J_{\aalpha}=0$)
\begin{equation}
\label{defab}
\dot J-\dot J_\text{w}=\dot J_\text{dw}(R_\text{out}).
\end{equation}
In this case the outer disk radius $R_\text{out}$ is equal to the 
radius above which all material is carried away by the disk 
wind. The corresponding disk wind mass loss rate $\dot M_\text{dw}(R_\text{out})$
is then given by Eqs.~\eqref{dmdtjed}, \eqref{paproxte}.
Note however that the formulae discussed in Sect.~\ref{secabla} are strictly 
valid only in the optically thick part of the disk (see Sect.~\ref{apdisopdep}).

If the net angular momentum loss rate $\dot J-\dot J_\text{w}$ is larger than
the maximum one, $\dot J-\dot J_\text{w}>\dot J_\text{dw}(\infty)$, then the
disk will be only partly ablated by the radiation. The net angular momentum
loss $\dot J-\dot J_\text{w}$ is in this case the sum of the parts carried by
the disk and disk wind, 
\begin{equation}
\label{partj}
\dot J-\dot J_\text{w}=\dot J_\text{dw}(\infty)+\dot J_{\aalpha}(\dot M_\text{d}),
\end{equation}
where $\dot J_{\aalpha}(\dot M_\text{d})$ is given by Eq.~\eqref{mamol}. Here one
can assume a conservative estimate of the isothermal disk with $p=0$.
From Eq.~\eqref{partj} the mass-loss rate carried away purely by the disk
$\dot M_\text{d}$
can be calculated, giving the total required 
mass-loss rate
as a sum of parts carried finally by the stellar wind ($\dot M_\text{w}$), disk
wind ($\dot M_\text{dw}$), and purely by the disk ($\dot M_\text{d}$) as
\begin{equation}
\label{partm}
\dot M=\dot M_\text{w}+\dot M_\text{dw}(\infty)+\dot M_\text{d}.
\end{equation}

The calculation of the functions $\dot J_\text{dw}(r)$ and $\dot M_\text{dw}(r)$
requires the knowledge of the line force parameters $\bar Q$ and $\alpha$. As
the NLTE calculation of these parameters for the disk wind environment are not
available, one can use their values derived for line driven winds for solar
metallicity, i.e., $\bar Q\approx2000$, and $\alpha\approx0.6$
\citep{gayley,pusle,nlteii}. For the metallicities other than the solar one the
scaling $\bar Q\sim Z$ can be used (here $Z$ is the mass fraction of heavier
elements), which is in a good agreement with the results of NLTE wind models
\citep{vikolamet,nlteii}.

\section{Other processes that may influence the outer disk radius}
\label{secot}

In addition to the radiative force, there may be other processes that may
influence the outer disk radius and consequently determine the required
mass-loss rate for a given angular momentum loss rate. For example, in binaries
the outer disk edge may be naturally truncated due to the presence of the
companion. However, the most uncertain part of the proposed model is connected
with the mechanism of the viscous transport, which may also influence the outer
disk radius.

\subsection{Loss of viscous coupling}

The magnetorotational instability \citep{barbus} is a promising mechanism to
explain the source of anomalous viscosity in accretion disks. As the dynamics of
accretion and decretion disks is similar, it is likely to be important also for
the angular momentum transfer in decretion disk. However the numerical
simulations of magnetorotational instability \citep[e.g.][]{skala,kralik}
concentrate on the inner parts of the disk, whereas the evolution close to the
sonic point is, to our knowledge, not very well studied. The stability
condition of the positive derivative of the angular frequency \citep{barbus}
$\de \Omega^2/\de r\geq0$
is fulfilled even in the supersonic wind region. The fact that the ratio of the
viscous timescale $\tau_\text{visc}\approx\zav{\alpha\Omega\zav{H/r}^2}^{-1}$ to
the growth timescale of the magnetorotational instability
$\tau_\text{MRI}\approx1/\Omega$ decreases with radius as
$\tau_\text{visc}/\tau_\text{MRI}\approx\zav{v_\text{K}/a}^2/\alpha$
\citep{kima}
indicates that in the outer parts of the disk where
the 
azimuthal
velocity 
is lower than the thermal speed
the magnetorotational instability would not be
effective. As this happens at supersonic velocities, this again supports our
conclusion that Eq.~\eqref{mamol} indeed gives the upper limit for the angular
momentum loss.

Moreover, the ratio of the particle kinetic energy to the absolute value of its
gravitational potential energy is roughly equal to $1/4$ at the critical point.
Consequently, for radius 
few times
larger than the critical one the disk material
may freely escape the star and the viscous support is no longer needed.

The loss of the viscous coupling may occur even before the radial disk expansion
becomes supersonic. In such a case for $\aalpha\rightarrow0$ from
Eq.~\eqref{momint} follows that the disk starts to be momentum conserving and
the location of the point where this occurs sets the outer disk radius
$R_\text{out}$.

Note also that the disk equations were derived assuming that the disk is
geometrically thin, i.e., $H\ll r$. However, at the critical point the ratio of
the disk scale height to radius $H/r\approx\sqrt{\frac{3}{10}}$ is of the order
of unity
(assuming isothermal disk).
Consequently, above the critical point
the vertical averaging used for the obtaining of
Eqs.~\eqref{kontrov}--\eqref{phirov} is no longer applicable.
On the other hand, because the angular momentum loss reaches a plateau below
this point, this effect has not a significant influence on our results.

\subsection{Influence of stars in a close neighbourhood}

For members of binaries or for stars in a very dense star cluster the disk can
be potentially truncated due to the influence of a nearby star. 

The nearby star could disrupt the disk by its gravitational interaction with the
disk. In this case the outer disk radius is that at which the gravitational
field of the nearby star starts to dominate, i.e., the Roche lobe radius in
the case of binaries.

If the disrupting star is luminous one, then it may disrupt the disk via the
radiative force. This case is analogous to the case of the radiative ablation
due to the central star. Consequently, we conclude that this effect would be
important only if the nearby star is located within a few radii from the
central star.

Finally, the nearby star may heat the disk material increasing the local sound
speed, and consequently decreasing the critical radius above which the disk
material may leave the star.

Taking all discussed disruption mechanisms together, we conclude that in the
case of the nearby star with a similar spectral type the disruption is effective
only if the disrupting star is at the distance lower or comparable to the
critical radius. If the nearby companion is able to
disrupt the disk, the angular momentum loss becomes less efficient, and the star
has to lose a larger amount of mass to keep the rotation velocity below the
critical one. Consequently, we expect larger disk mass-loss in close binaries
and in very dense star clusters.

\subsection{The disk build-up and its angular momentum}

In the analysis presented here we used an assumption of constant required
angular momentum loss rate, which enabled us to use stationary equations. This
assumption is reasonable in most phases of the stellar evolution, as the
evolutionary timescale is much longer than the typical timescale of the disk
build-up, which is of the order of years \citep{okaform,josip}. This also means
that the transitional processes that occur when the star reaches or leaves the
critical limit are more complicated than studied here.

In the course of the stellar evolution, when the surface rotational velocity
reaches the critical limit, in a first time the disk appears because it is
feeded by the mechanical mass loss. The disc grows and part of it is ablated and
part is transported away via viscous coupling until an equilibrium between
the required angular momentum loss rate and mass-loss rate is achieved. During
this process the disk own angular momentum could be of some importance.

On the other hand, when the star leaves the critical limit, an inner part of the
remaining disk is accreted on the star while other parts are expelled into the
interstellar medium \citep{okaform}.


\subsection{Implication for stars with disk}

The processes discussed here might be relevant also for other stars with disks.
For example, the disk radiative ablation might be one of the reasons why the Be
phenomenon is typical for B stars only, whereas for more luminous O stars any
disk could be destroyed by the radiative force.

Similar effects should also be present in accretion disks during star formation.
In more luminous stars the radiative ablation could contribute to the disk
photoevaporation \citep[e.g.,][]{adams,alexander} in dispersing of the disk.
Moreover, a similar process of the angular momentum transfer is present also in
the accretion disks of these stars, consequently influencing the distribution of
the rotational speeds on the ZAMS.

\subsection{Implications for first stars}

Mechanical mass loss through a decretion disk can be a ubiquitous phenomenon
especially for Pop III or very metal poor stars. Indeed as shown by
\citet{ekmemaba} pure hydrogen-helium  Pop III stars with masses above
60~M$_\odot$, beginning their evolution on the ZAMS with a surface velocity
around 70\% of the critical angular velocity, will reach the critical velocity
during the MS phase. This arises because of two effects: first angular momentum
is transported from the inner regions to the surface during the MS phase;
second, the angular momentum accumulates at the surface since it is not removed
by stellar winds. Note that in the absence of metals hydrogen and helium
are unable to drive a line-driven wind being nearly completely ionized
\citep{bezvi}.


As hydrogen-helium first stars are unable to launch a line-driven wind, we
expect the radiative ablation to be inefficient close to the star. On the other
hand, at larger distances a nonnegligible fraction of hydrogen could become
neutral, enabling the possibility of disk radiative ablation.

The disk wind mass-loss rate in such case could be described as a flow with a
very low value of $\bar Q$ (corresponding likely just to Ly$\alpha$ line force).
A rough estimate of the disk wind mass-loss rate in this case could be obtained
inserting instead of $\dot M_\text{CAK}$ the single line mass-loss rate estimate
$\dot M\approx L/c^2$ \citep{lusol} in Eq.~\eqref{dmdtjed}. Anyway, in most
cases such flow would be likely inefficient, especially because the disk wind
mass-loss rate originates close to the star (see Fig.~\ref{palfaobr}).
Consequently, the relation between the mass-loss rate required for a given
angular momentum loss rate would be given by the wind-free condition
Eq.~\eqref{mamol}.

\subsection{Future work}

The most uncertain ingredients of a proposed model are the viscous coupling, the
disk temperature distribution and the radiative ablation. To include these
processes we applied the same description used in the accretion disk theory and
the theory of radiatively driven winds of hot stars. This may not be completely
adequate for the description of decretion disk especially at large distances
from the star studied here. Consequently, future work should address these
problems.

\section{Conclusions}
\label{seccon}

We examine the mechanism of the mass and angular momentum loss via decretion
disks associated with near-critical rotation. The disk mass loss is set by the
angular momentum needed to keep the stellar rotation at or below the critical
rate. We study the potentially important role of viscous coupling in outward
angular momentum transport in the decretion disk, emphasizing that the specific
angular momentum at the outer edge of the disk can be much larger than at the
stellar surface. For a given stellar interior angular momentum excess,  the mass
loss required from a decretion disk can be significantly less than invoked in
previous models assuming a direct, near-surface release.

The efficiency of the angular momentum loss via disk depends on the radius at
which the viscous coupling ceases the transport the angular momentum to the
outflowing material. When the radiative force is negligible, we argue that this
likely happens close to the disk sonic (critical) point setting the most
efficient angular momentum loss. In the opposite case, when the radiative force
is nonnegligible,
there is not a single point beyond which the viscous coupling disappears. The
disk is continuously ablated below the sonic point, and the ablated material
ceases to be viscously coupled, decreasing the efficiency of angular momentum
loss.

We describe the method to include these processes into evolutionary
calculations.
The procedure provided enables to calculate the mass-loss rate
necessary for a required angular momentum loss just from the stellar and line
force parameters. 
We can distinguish three different physical circumstances:
\addtolength\leftmargini{8mm}
\begin{itemize}
\item[case A:]
When the disk wind is able to remove the whole excess of angular
momentum (the disk is completely ablated by the wind, see Eq.~\eqref{defab}) then
the outer disk radius is given by Eq.~\eqref{defab}, and the required mass loss is given
by Eq.~\eqref{dmdtjed}. The limiting case
$R_\text{out}\approx R_\text{eq}$ would then correspond to the near surface
release of the matter without any disk.
Note that in a rare case when the analysis leads to
$R_\text{out}>R_\text{crit}$ the radius $R_\text{crit}$ should be used as the
outer disk radius (case B).
The expressions presented in the paper are given in
the hypothesis of an optically thick
disk and should be appropriately modified for optically thin disks.
\item[case B:]
If the raditiave force is not able to remove sufficient angular
momentum (the disk is not completely ablated) then part of the excess
angular momentum must be carried away by the disk (Eq.~\eqref{partj}).
In this case Eqs.~\eqref{partj}, \eqref{partm}
can be used to estimate the mass-loss rate. The outer disk
edge
could be
identified with the critical point.
\item[case C:]
If the effects of the radiative force are negligible, then the
whole excess of angular momentum
is carried away by the disk  and the the outer disk edge is approximately
given by $R_\text{crit}$ and the required mass-loss rate could be derived from
Eq.~\eqref{mamol}.
\end{itemize}
\addtolength\leftmargini{-8mm}

Finally, we note that, in absence of strong magnetic field,
many of the features discussed here  may also be applicable to the case of 
star-formation accretion disks. 

\begin{acknowledgements}
This work was supported by grant GA\,\v{C}R 
205/08/0003.
This research made use of NASA's Astrophysics Data System.
\end{acknowledgements}

\newcommand{\actob}{Active OB-Stars:
        Laboratories for Stellar \& Circumstellar Physics,
        eds. S. \v{S}tefl, S. P. Owocki, \& A.~T.
        Okazaki (ASP, San Francisco), }

\appendix
\section{Disk wind mass-loss rate}
\label{secablaap}

\subsection{Disk optical depth}
\label{apdisopdep}

In the case when the disk is optically thick in continuum, the disk outflow may
be driven not only by the radiation from the stellar surface, but also by the
stellar radiation reprocessed by the disk \citep{abla}. To estimate the optical
depth of the disk, let us assume hydrogen and helium to be ionized in the disk.
In this case a significant part of the disk optical depth originates due to the
light scattering on free electrons (for wavelengths lower than that
corresponding to the Balmer or Lyman jump also bound-free transitions may
contribute). The transverse optical depth is then roughly given by
$\tau=\int\kappa_\text{e}\rho\,\de z=\kappa_\text{e}\Sigma$, where
$\kappa_\text{e}$ is the Thomson scattering cross-section per unit of mass. The
disk is optically thick in the vertical direction ($\tau>1$) if the mass-loss
rate is larger than
\begin{equation}
\dot M > \frac{2\pi r v_r}{\kappa_\text{e}}\approx
10^{-12}\,\text{M}_\Sun\,\text{year}^{-1}
\zav{\frac{r}{1\,\text{R}_\Sun}}
\zav{\frac{v_r}{1\,\text{m}\,\text{s}^{-1}}}.
\end{equation}
For a given mass-loss rate the disk is optically thick close to the star,
while becoming optically thin at larger distances. For example, for a typical
disk mass-loss rate required by the evolutionary calculations
$10^{-5}\,\text{M}_\Sun/\text{year}^{-1}$ \citep[e.g.][]{eznula} the disk is
optically thick even at large distances from the star
$r\approx10^3\text{R}_\Sun$ for subsonic radial velocities. Consequently, in
realistic situations the disk is likely to be optically thick, at least close to
the star, resembling the "pseudophotosphere" of Be stars \citep[e.g.][]{kouha}.

Contrary to very dense hot star winds (where the radiative flux comes from
regions below the photosphere), here we expect that the wind from the optically
thick disk starts to accelerate above the point where the disk optical depth is
unity. Numerical results show that the height of this point
is comparable to the disk scale height $H$ for
moderate disk mass-loss rates $\dot M\lesssim
10^{-5}\,\text{M}_\Sun/\text{year}^{-1}$. Consequently, we shall neglect the
disk geometrical height in in our analyze  here and assume that the disk wind
originates from the equatorial plane.

\subsection{Disk wind equations}

\begin{figure*}[t]
\centering
\includegraphics[width=0.90\hsize]{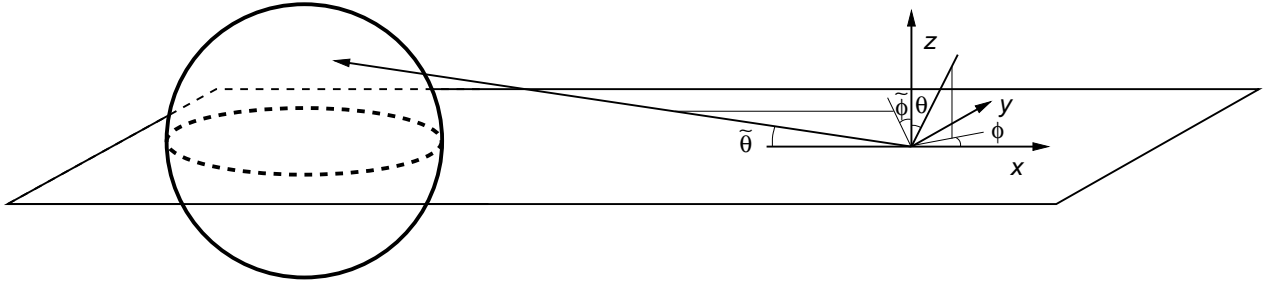}
\caption{The coordinate system for the calculation of the radiative force}
\label{coord}
\end{figure*}

The outflow from the
optically thick
disk irradiated by the central star can be understood
within the framework of the wind driven by external irradiation \citep{abla}. We
study the disk outflow in noninertial frame corotating with the disk. We use the
Cartesian coordinates with $z$ axis perpendicular to the disk
(see Fig.~\ref{coord}).
The disk wind
originates in the disk plane $z=0$. We assume purely vertical flow with velocity
$v_z(z)$ and we neglect a potentially important part of the radiative force due
to the Keplerian velocity gradient \citep{diskra}.

The stationary continuity equation 
\begin{equation}
\nabla \zav{\rho\mathbf{v}}=0,
\end{equation}
takes within our assumptions the form of
\begin{equation}
\label{kontrovvit}
\frac{\partial}{\partial z}\zav{\rho v_z}=0,\quad\text{or}\quad
\dot m\equiv\rho v_z=\text{const.},
\end{equation}
where $\dot m$ is the disk wind mass-loss rate per unit of disk surface.

The radiative force of the ensemble of lines in the Sobolev approximation is then
\citep{rybashumrem,nerad,gayley}
\begin{equation}
\label{zarsil}
\mathbf{g}_\text{rad}=
\frac{c^{-2\alpha}}{1-\alpha}\zav{\frac{\kappa_\text{e}\bar Q}{c}}^{1-\alpha}
\oint I(\mathbf{n})
\zav{\frac{\mathbf{n}\nabla \zav{\mathbf{n}\mathbf{v}}}{\rho}}^\alpha
\mathbf{n}\,\de\Omega,
\end{equation}
where $\bar Q$ and $\alpha$ are line force parameters. Ignoring the incoming
beam, and simply assuming all the locally normal incident radiation
from one hemisphere
is directly
reflected upward in vertical beam normal to the disk, the intensity is given by
\begin{multline}
I(\mu,\phi)=\delta(\phi)\delta(\mu-1)
\int_{-\frac{\pi}{2}}^{\frac{\pi}{2}}\de\tilde\phi\,\cos\tilde\phi
\int_{\mu_*}^1 \de\tilde\mu\sqrt{1-\tilde\mu^2}\,I_* \\*
=\frac{2}{\pi}\delta(\phi)\delta(\mu-1)F\frac{r^2}{R^2}\int_{\mu_*}^1
\sqrt{1-\tilde\mu^2}  \de\tilde\mu,
\end{multline}
where $I_*=\zav{r/R}^2F/\pi$ is the emergent intensity from the stellar
photosphere, $F$ is the radiative flux at radius $r$,
$R$ is the stellar radius,
$\tilde\mu=\cos\tilde\theta$, $\tilde\phi$ are spherical
coordinates with origin at the stellar centre, $\mu_*=\sqrt{1-R^2/r^2}$, and
$\mu$, and $\phi$ are the direction cosine and azimuthal angle measured from
the disk plane (see Fig.~\ref{coord}).
The $z$-component of the radiative force in this case is
\citep[Eq.~\eqref{zarsil},][]{abla}
\begin{equation}
g_\text{rad}=C\azav{v_z\frac{\partial v_z}{\partial z}}^\alpha f_z,
\end{equation}
where
\begin{equation}
\label{fzil}
f_z=\frac{2}{\pi}F\frac{r^2}{R^2}\int_{\mu_*}^1\sqrt{1-\tilde\mu^2}\,
\de\tilde\mu=
\frac{F}{\pi}\frac{r^2}{R^2} \arccos(\mu_*) - \frac{F}{\pi}\frac{r}{R} \mu_*,
\end{equation}
and
\begin{equation}
\label{blansko}
C=\frac{1}{1-\alpha}\zav{\frac{\kappa_\text{e}\bar Q}{c}}^{1-\alpha}
\zav{\dot m c^2}^{-\alpha}.
\end{equation}

The $z$-component of the momentum equation including the gravity term and
neglecting the gas pressure term is
\begin{equation}
\label{zmom}
v_z\frac{\partial v_z}{\partial z}=
C\azav{v_z\frac{\partial v_z}{\partial z}}^\alpha
f_z-\frac{GMz}{\zav{r^2+z^2}^{3/2}}.
\end{equation}
The vertical momentum equation Eq.~\eqref{zmom} can be solved using the
transformations
\begin{subequations}
\label{trans}
\begin{align}
w&=\frac{r}{2GM}v_z^2,\\
\zeta&=\frac{z}{r},\\
K&=Cf_z\zav{\frac{GM}{r^2}}^{\alpha-1},
\end{align}
\end{subequations}
yielding
\begin{equation}
\label{wrovskal}
w'=Kw'^\alpha-\frac{\zeta}{\zav{1+\zeta^2}^{3/2}},
\end{equation}
where the prime denotes the derivative with respect to $\zeta$. This equation has a
critical point 
\begin{equation}
\label{crit}
1-\alpha K_\text{c} w_\text{c}'^{\alpha-1}=0,
\end{equation}
where the subscript c denotes the critical point values,
from which using Eq.~\eqref{wrovskal} we derive
\begin{equation}
\label{wcprime}
w_\text{c}'=\frac{\alpha}{1-\alpha}\frac{\zeta_\text{c}}{\zav{1+
\zeta_\text{c}^2}^{3/2}}.
\end{equation}
The location of the critical point above the disk plane can be derived from the
regularity condition (CAK), which yields that the critical point occurs at the
point of maximum of $z$ component of the gravity acceleration at a given
streamline,
\begin{equation}
\label{zetac}
\zeta_\text{c}=\frac{1}{\sqrt{2}}.
\end{equation}
Hence, the point of the maximum acceleration acts as the throat of the nozzle
flow \citep{feshlo}. 

The total disk wind mass-loss rate is then given by an integral of the mass-loss
rate per unit of the disk surface $\dot m$ between the equatorial radius
$R_\text{eq}$ and the outer disk radius $R_\text{out}$
\begin{equation}
\label{dmdtdef}
\dot M_\text{dw}(R_\text{out})=2\times2\pi
\int_{R_\text{eq}}^{R_\text{out}}\dot m r\,\de r,
\end{equation}
where from Eqs.~\eqref{blansko}, \eqref{trans}, \eqref{crit}
\begin{equation}
\label{malem}
\dot m=\frac{1}{c^2}
\zav{\frac{\kappa_\text{e}\bar Q}{cw_\text{c}'}}^{\frac{1-\alpha}{\alpha}}
\zav{\frac{f_z\alpha}{1-\alpha}}^{\frac{1}{\alpha}}
\zav{\frac{GM}{r^2}}^{\frac{\alpha-1}{\alpha}}.
\end{equation}
The factor of 2 in Eq.~\eqref{dmdtdef} comes from the fact that the wind
originates from both sides of the disk. Consequently, the total disk wind
mass-loss rate is
\begin{equation}
\label{mdotdw}
\dot M_\text{dw}(R_\text{out})= {\frac{\alpha}{1-\alpha}}\frac{L}{c^2}
\zav{\Gamma\bar Q}^{\frac{1-\alpha}{\alpha}}P_1\zav{\frac{R_\text{out}}{R}},
\end{equation}
where the Eddington parameter $\Gamma=\kappa_\text{e} L /\zav{4\pi GMc}$, and
\begin{equation}
\label{palfa}
P_\ell(x_\text{out})=
\zav{\frac{\alpha}{1-\alpha}}^{\frac{1-\alpha}{\alpha}}
\int_{3/2}^{x_\text{out}} w_\text{c}'^{\frac{\alpha-1}{\alpha}}
\zav{\frac{f_z}{F}}^{\frac{1}{\alpha}} \frac{\de x}{x^\ell},
\end{equation}
and $F$ is the flux at radius $r$. Comparing with the CAK mass-loss rate
estimate Eq.~\eqref{mcak}
we have
\begin{equation}
\label{mdotdwcak}
\dot M_\text{dw}(R_\text{out})=
P_1\zav{\frac{R_\text{out}}{R}}\,\dot M_\text{CAK}.
\end{equation}

\begin{figure}
\centering
\includegraphics[width=0.9\hsize]{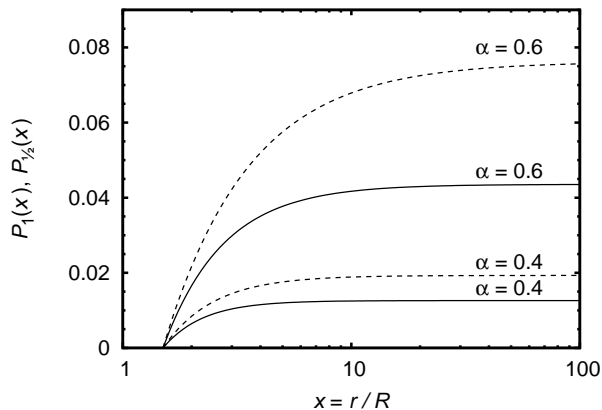}
\caption{The behaviour of $P_1(x)$ (solid line), and $P_{\frac{1}{2}}(x)$
(dashed line, Eq.~\eqref{palfa}) for different values of CAK parameter
$\alpha$.}
\label{palfaobr}
\end{figure}

Assuming the disk wind is not viscously coupled to the disk, the total angular
momentum loss via the disk wind is 
\begin{equation}
\label{job}
\dot J_\text{dw}(R_\text{out})=2\times2\pi\int_{R_\text{eq}}^{R_\text{out}}
r^2v_\phi\dot m\, \de r,
\end{equation}
where $\dot m$ is given by Eq.~\eqref{malem}. As the disk wind originates mainly
from the regions close to the star (with $r/R\lesssim10$, see
Fig.~\ref{palfaobr}), where the
azimuthal
velocity is roughly equal to the Keplerian one (see
Fig.~\ref{500_2_a}), we can assume $v_\phi\approx v_\text{K}(r)$ in
Eq.~\eqref{job} and consequently
\begin{equation}
\label{jdotdw}
\dot J_\text{dw}(R_\text{out})= \frac{\alpha}{1-\alpha}\frac{L}{c^2}
\zav{\Gamma\bar Q}^{\frac{1-\alpha}{\alpha}}R\,v_\text{K}(R)\,
P_{\frac{1}{2}}\zav{\frac{R_\text{out}}{R}}.
\end{equation}
Again, using the CAK mass-loss rate estimate Eq.~\eqref{mcak} the disk wind
angular momentum loss
\begin{equation}
\dot J_\text{dw}(R_\text{out})=R\,v_\text{K}(R)\,P_{\frac{1}{2}}
\zav{\frac{R_\text{out}}{R}}\dot M_\text{CAK}
\end{equation}
is by a factor of $P_{\frac{1}{2}}(R_\text{out}/R)$ larger than the angular
momentum loss due to the CAK wind launched from equator of hypothetical
critically rotating spherical star with radius $R$.

Expansion of Eq.~\eqref{fzil} about $R/r=0$ shows that far away from the star,
$f_z/F \approx 2R/(3\pi r)$. The comparison between a precise formula
Eq.~\eqref{fzil} and its asymptotic form shows apart from a small region
$1<x<3/2$, the agreement is actually quite good. If we use this asymptotic form
over the full range from $x=3/2$ to $x_\text{out}$ in Eq.~\eqref{palfa}
we find (using Eqs.~\eqref{wcprime}, \eqref{zetac})
\begin{equation}
\label{paprox}
P_\ell(x_\text{out})=\frac{2\pi^{-\frac{1}{\alpha}}
3^{\frac{1}{2\alpha}-\frac{3}{2}}}{\frac{1}{\alpha}+\ell-1}
\hzav{\zav{\frac{3}{2}}^{1-\ell-\frac{1}{\alpha}}-x_\text{out}^{1-\ell-\frac{1}{\alpha}}}.
\end{equation}
Our analytical results (see Fig.~\ref{palfaobr}) are in agreement with numerical
calculations of \citet{prostod} that show the disk mass loss is dominated
by material arising from the inner region of the disk ($r<10\,R$). Consequently,
if the radiative force is not strong enough to disrupt the disk close to the
star, it is unlikely that it would be able to do so in the outer parts of the
disk.

The modern hot star wind models give slightly lower estimate of the mass-loss
rate than the CAK formula Eq.~\eqref{mcak} due to inclusion of finite disk
correction. However, because formula
Eq.~\eqref{paprox} for a typical value of $\alpha\approx0.6$
\citep{pusle,nlteii} gives for $x_\text{out}\rightarrow\infty$ the value of
$P_1\approx0.04$, the disk wind mass-loss rate is significantly lower than the
stellar wind mass-loss rate. Similarly, the value of $P_{1/2}=0.08$ indicates
that the angular momentum loss from the disk wind is also lower than the
angular momentum loss due to the stellar wind.

\end{document}